\documentclass[prd,preprint,nofootinbib]{revtex4}

\usepackage{amsmath,amssymb}
\usepackage{graphicx}
\usepackage{dcolumn}
\usepackage{bm}
\usepackage{cancel}

\begin{document}
\title{Bulk Fermions in Warped Models with a Soft Wall}

\date{\today}

\author{S. Mert Aybat}
\author{Jos\'e Santiago}
\affiliation{Institute for Theoretical Physics, ETH, CH-8093,
  Z\"urich, Switzerland}

\begin{abstract}

We study bulk fermions in models with warped extra dimensions in the
presence of a soft wall. 
Fermions can acquire a position dependent bulk Dirac mass
that shields them from the deep infrared, allowing for a systematic
expansion in which electroweak symmetry breaking effects are treated
perturbatively. 
Using this expansion, we analyze properties of bulk fermions in the
soft wall background. These properties include the realization of
non-trivial  boundary conditions that simulate the ones commonly used
in hard wall models, the analysis of the flavor structure of the model
and the implications of a
heavy top. We implement a soft wall model of electroweak symmetry
breaking with custodial symmetry and fermions propagating in the bulk.
We find a lower bound on the masses of the first bosonic resonances,
after including the effects of the top sector on electroweak precision
observables for the first time, 
of $m_{KK} \gtrsim 1 -3$ TeV at the $95 \%$ C.L.,
depending on the details of the Higgs, and
discuss the implications of our results for LHC phenomenology.  

\end{abstract}

\pacs{}

\maketitle

\section{Introduction}

Models with warped extra dimensions~\cite{Randall:1999vf}
offer a rich new avenue
towards our understanding of the stability of the electroweak (EW) scale
against ultraviolet physics. 
The AdS/CFT
correspondence~\cite{Maldacena:1997re}
applied to models compactified on a slice of
AdS$_5$~\cite{ArkaniHamed:2000ds},  
resulted in an AdS$_5$/CFT dictionary relating the 5D models to their
dual 4D strongly coupled field theories.
The intensive effort put into the study of models with warped extra
dimensions has resulted in a number of realistic models of natural
electroweak symmetry breaking (EWSB) based on
them~\cite{Carena:2006bn}.~\footnote{See
also~\cite{Cacciapaglia:2006gp} for Higgsless models based on the same
idea.}
Custodial symmetry~\cite{Agashe:2003zs} 
and a custodial protection of the
$Zb\bar{b}$ coupling~\cite{Agashe:2006at} have proven essential to get
realistic models with light enough new particles to make them
accessible at the LHC.~\footnote{Alternative constructions that do not
use these custodial protections have been explored
in~\cite{Davoudiasl:2002ua}.}
Surprisingly, the immense majority of the models follow the original
in the choice of background, AdS$_5$, and the way the conformal
symmetry is spontaneously broken, by suddenly ending the extra
dimension in what is called the infrared (IR) brane. 
This hard wall realization
of the spontaneous breaking of the conformal invariance corresponds,
in the dual 4D picture to a breaking by 
an operator of infinite scaling dimension.
Instead, one could replace the IR brane with a soft wall so that the extra
dimension expands to infinity but there is a departure from
conformality in the IR either by a modification of the metric or by
the introduction of a dilaton. This corresponds in the dual picture to
conformal breaking by an operator of finite dimension.

Soft wall models were introduced in~\cite{Karch:2006pv} 
to model the observed
Regge trajectories in hadronic resonances through
five-dimensional duals (hard wall models in AdS$_5$ give a mass
scaling $m_n^2 \sim n^2$ instead of the observed $m_n^2 \sim n$). 
Very recently, soft wall realizations of models of EWSB
were presented
in~\cite{Falkowski:2008fz,Batell:2008me}.
The spectrum of new resonances is affected by the soft wall, with
behaviors that range from a (modified) discrete spectrum to a
continuous spectrum with or without a mass gap. This provides new
realizations of hidden-valley like models~\cite{Strassler:2006im} 
and unparticle
physics~\cite{Georgi:2007ek}
(see~\cite{Cacciapaglia:2008ns,Falkowski:2008yr}).   
An analysis of the bosonic sector of these models shows that the
constraints on the masses of new particles
from the $S$ parameter alone can be relaxed with respect to hard wall
models and new bosonic resonances as light as $1$ TeV, or even
lighter, are compatible with current limits on the $S$
parameter~\cite{Falkowski:2008fz,Batell:2008me}. Earlier
attempts~\cite{Shiu:2007tn} used a smooth deviation
from AdS in the IR given by a power law, instead of the exponential
form we will use, and considered the Higgs to be localized at a fixed
position in the bulk (on a non-gravitating brane). 
The result was that the bound on the mass of the lightest
Kaluza-Klein (KK) 
modes did not appreciably change with respect to the case of a
hard wall in the family of parameterizations used.

The
fermionic sector, however, was not studied in detail in either of
these works and was not included in the analysis of the
EW constraints. This is fine in general for the lighter fermions, which
can to a good approximation be considered as fields living in the UV
brane. The large mass of the top quark, however, does not allow us to
neglect the fact that third generation quarks have to propagate in the
bulk. This results, in hard wall models, in an important effect of the
top sector in EW precision tests through their contribution
mainly to the $T$ parameter and the $Zb\bar{b}$ coupling at the loop
level. The aim of this work is to systematically study bulk fermions
in soft wall models. This allows us to discuss the impact of the top
sector on EW observables and therefore obtain reliable bounds
on the mass of the new resonances. Also, although 
considering light fermions as bulk
fields is irrelevant regarding EW observables, it is not
regarding the flavor structure of the model. Out
formalism provides the tools to analyze the flavor constraints
in soft wall models.

The outline of the paper is as follows. In section~\ref{fermions} we
consider bulk fermions in warped extra dimensions with a soft wall. 
We develop tools that
allow us to study the implications of bulk fermions, from the light
generations to the top quark, in EW precision tests and
flavor physics. In section~\ref{ew:constraints} we use these tools to
analyze the EW constraints in custodial models with a soft
wall, including for the first time the effect of the top sector in
such models. Section~\ref{conclusions} is devoted to a discussion of
our results, including the distinctive collider implications of
soft wall models, and to our conclusions.
We have included in Appendices~\ref{background}
and~\ref{bosons} technical details on the background 
and the expansion of bosonic fields for completeness.

\section{Bulk Fermions in the Soft Wall \label{fermions}}

A soft wall can be implemented in models with warped extra dimensions
either through the position dependent vacuum expectation value (vev)
of a dilaton field or by a
modification of the metric in the IR. We choose the former
approach and work in an AdS$_5$ background,
\begin{equation}
ds^2=a(z)^2 ( \eta_{\mu\nu} dx^\mu dx^\nu - dz^2),\label{metric}
\end{equation}
with $\mu,\nu=0,\ldots,3$ and $a(z)=\frac{L_0}{z}$. $z\geq L_0$ is the
coordinate along the extra dimension and $L_0 \approx
M_{\mathrm{Planck}}^{-1}$ is the inverse AdS$_5$ curvature scale. The
end-point at $z=L_0$ is commonly denoted as the ultraviolet (UV) brane.
The soft wall is generated by the $z$-dependent vev of a dilaton
field $\Phi(z)$. The matter action reads
\begin{equation}
S_{\mathrm{matter}}=\int d^5x \sqrt{g} e^{-\Phi}
\mathcal{L}_{\mathrm{matter}}. \label{action}
\end{equation}
The spectrum of bosonic fields is sensitive to the dilaton profile.
In particular, depending on its behavior at large $z$, it
can lead to a discrete or continuous spectrum (with or without a mass
gap). In the case of a discrete spectrum, the inter-mode spacing
also depends on the dilaton profile. In this article we will focus on
a quadratic dilaton profile
\begin{equation}
\Phi(z)=\left(\frac{z}{L_1}\right)^2,
\end{equation}
which gives rise to a discrete spectrum with masses scaling as $m_n^2
\sim n$ as opposed to the usual $m_n^2 \sim n^2$ found in hard wall
models. Note that 
$L_1$ here is not the position of any brane 
but the scale at which the effect of the dilaton
becomes sizable and the solution departs from standard AdS.
Still, $L_1^{-1}$ determines the mass gap and 
if we want our soft wall model to solve the hierarchy problem,
we should have $L_1^{-1} \sim$ TeV. 
This background can be obtained dynamically as a solution to Einstein
equations in the presence of an extra bulk tachyonic
scalar, as was shown in~\cite{Batell:2008me}. The solution for bulk
bosonic fields were analyzed in detail in that reference and we just
collect the relevant results in the appendices. Bulk fermions were
also discussed in~\cite{Batell:2008me} but they are somewhat
problematic in soft wall models and not many details could be
investigated. The main result of this article is to develop the tools
needed to analyze in great generality the phenomenological implications of
bulk fermions in soft wall models.

\subsection{The problem with fermions in the soft wall}
Consider a five-dimensional fermion, $\Psi(x,z)$, 
in our soft wall background. Its action
reads
\begin{equation}
S=\int d^5x \, \sqrt{g} e^{-\Phi} \Big[ \frac{1}{2}\big( 
\bar{\Psi} e^N_A \Gamma^A i D_N \Psi_L 
- i D_N \bar{\Psi} e^N_A \Gamma^A \Psi  \big)
- M \bar{\Psi}\Psi\Big],
\end{equation}
where $N,A=\{\mu, 5\}$ run over five dimensions in the curved and 
the tangent spaces respectively, $e^N_A$ is the f\"unfbein,
$D_N$ is the gravitationally covariant derivative, $\Gamma^A = \{
\gamma^\mu , -i \gamma^5 \}$ are the Dirac matrices in 5 dimensions 
and $M$ is a bulk
Dirac mass that is left unspecified for the moment.
After integration by parts~\footnote{Integration by parts will result
  in boundary terms that have to be properly taken into account. We
  assume that the required boundary terms exist to make our choice of
  boundary conditions compatible with the variational principle.}
the action can be written as
\begin{equation}
S=
\int d^5x \, a^4 e^{-\Phi} \bar{\Psi} \left[
i \cancel{\partial} +\left( \partial_5 + 2 \frac{a^\prime}{a} - \frac{1}{2}
\Phi^\prime \right) \gamma^5 - a M \right] \Psi 
=\int d^5x \, \bar{\psi} \big[
i \cancel{\partial} + \partial_5 \gamma^5 - a M \big] \psi, 
\label{bulk:fermion:action}
\end{equation}
where in the second equation we have defined
\begin{equation}
\psi(x,z)\equiv a^{2}(z) e^{-\Phi(z)/2} \Psi(x,z).
\end{equation}
The action written in terms of the field $\psi$ shows the main
problem. The dilaton and the metric have
disappeared from the action, except for the factor of the
metric multiplying the Dirac mass. 
This latter term is actually very important. In the absence of a bulk
mass, even in hard wall models, fermions would know nothing about the
background and their KK expansion would be in terms of trigonometric
instead of Bessel functions. A constant bulk Dirac mass allows the
fermions to have a KK expansion more according to the AdS$_5$
background and for a particular value $M=1/(2L_0)$ they behave
exactly the same as bulk gauge bosons in the same
background. Unfortunately this is not enough in soft wall models. 
If we assume the standard choice of
a constant bulk Dirac mass, then the corresponding zero mode, when
allowed by the boundary conditions, does not go to zero rapidly enough
at large $z$. Even when it is normalizable, it will invariably 
have strong coupling with some gauge boson KK modes. The situation
would not improve if we
modeled the soft wall with a modified metric instead of a dilaton. In
that case, the modification of the metric goes in the direction of 
making it go to zero faster than $z^{-2}$ (for instance exponentially) 
at large $z$. 
The effect of the bulk mass is then even less 
important and the zero mode is simply not normalizable. 
The authors of~\cite{Batell:2008me} noted
that, if the Higgs has a $z$-dependent profile that grows towards the
IR, as would be expected in a model that solves the hierarchy problem, 
its Yukawa coupling to fermions induces a bulk mass term that
shields the fermions from the deep IR and can thus solve this problem. 
Unfortunately this solution is difficult to implement in
practice, since the Yukawa coupling gives a $z$-dependent mass that
mixes two different 5D fermions. 
Even neglecting inter-generational mixing, this $z$-dependent mixing
makes it impossible to get analytic solutions except in the very
particular case of common bulk Dirac mass for the two five-dimensional
fermion fields coupled through the Higgs. This idea has been further
explored recently, for common bulk Dirac masses and different values of the
Higgs profile in~\cite{Delgado:2009xb}.

\subsection{Our solution}

A full analysis of the implications of bulk fermions in EW
and flavor physics requires, however, to go beyond the
particular case of common bulk Dirac masses. 
Recall that the only link that bulk fermions have to the background
they live in comes from the mass term. A constant mass term gives
information about the metric but it does not about the dilaton. A
natural assumption is therefore to consider that, besides the
constant bulk Dirac mass, bulk fermions acquire a $z$-dependent mass
that comes from a direct coupling to the dilaton. 
This mass provides the the missing link with the soft wall, shielding
the fermions from the deep IR without the need of the Yukawa coupling. 
In the most general case, this mass can mix different generations,
which would make the coupled first order equation impossible to solve
analytically. However, we can neglect inter-generational mixing in the
bulk Dirac masses and introduce it through Yukawa couplings.
Provided we do not introduce too large 5D Yukawa couplings
we can treat EWSB perturbatively in which case the inter-generational
mixing given by the Yukawa couplings does not represent any technical
problem.  

Our starting
point is the action in Eq.~(\ref{bulk:fermion:action}) with a bulk Dirac
mass given by
\begin{equation}
M(z)=\frac{c_0}{L_0} + \frac{c_1}{L_0} \frac{z^2}{L_1^2},
\label{bulk:mass}
\end{equation}
where $c_{0,1}$ are dimensionless constants expected to be order one.
The equations of motion derived from the fermionic action read
\begin{equation}
i\cancel{\partial} \psi_{L,R} +(\pm \partial_5-aM)\psi_{R,L}=0,
\end{equation}
where $L$ and $R$ stand for the left-handed (LH) and right-handed (RH)
components
respectively, $\gamma^5
\psi_{L,R}=\mp \psi_{L,R}$.
A standard expansion in KK modes,
\begin{equation}
\psi_{L,R}(x,z) = \sum_n f_n^{L,R}(z) \psi^{(n)}_{L,R}(x),
\end{equation}
with $i \cancel{\partial} \psi^{(n)}_{L,R}(x) = m_n
\psi^{(n)}_{R,L}(x)$ gives the equations for the fermionic profiles
\begin{equation}
(\partial_5 \pm a M)f_n^{L,R} = \pm m_n f_n^{R,L}.
\end{equation}
The orthonormality condition
\begin{equation}
\int_{L_0}^\infty f_n^L f_m^L = \int_{L_0}^\infty f_n^R f_m^R = 
\delta_{nm},
\end{equation}
then gives the action as a sum over four-dimensional Dirac KK modes and
 possibly massless zero modes,
\begin{equation}
S=\int d^4x \, \sum_n \bar{\psi}^{(n)} [ i \cancel{\partial} - m_n ]
\psi^{(n)}.
\end{equation}
The first order coupled equations for the fermionic profiles can 
be iterated to give two decoupled second order differential equations 
\begin{equation}
\Big[\partial_5^2 \pm (a M)^\prime -(aM)^2 +m_n^2\Big] f_n^{L,R}(z) =0.
\end{equation}
Inserting the expression of the metric and the mass, we get for the LH
profile,
\begin{equation}
\left[
\partial_5^2 - \frac{c_0(c_0+1)}{z^2} + \frac{c_1}{L_1^2} (1-2c_0) +
m_n^2 -\frac{c_1^2 z^2}{L_1^4} \right] f_n^L = 0\,,
\end{equation}
while the RH solution is identical to the LH one with the identification
$c_{0,1} \to -c_{0,1}$.
This equation can be put in the form of Kummer's equation
\begin{equation}
\big[x \partial_x^2 +(b-x)\partial_x -a \big] g(x) =0, 
\end{equation}
by means of the following changes of variables 
\begin{equation}
f(z) = e^{-\frac{|c_1|z^2}{2L_1^2}} z^{-c_0} g(z), 
\quad
x=\frac{|c_1|z^2}{L_1^2}, \label{fermion:change}
\end{equation}
where
\begin{equation}
a=\frac{1-2c_0}{4}-\frac{c_1(1-2c_0)+L_1^2m_n^2}{4 |c_1|},
\quad
b=\frac{1}{2}-c_0.
\end{equation}
The normalizable solutions of the coupled linear equations can then be
written as,
\begin{eqnarray}
&& \left . 
\begin{array}{l}
f_n^L(z)=
N_n z^{-c_0} e^{-\frac{c_1z^2}{2L_1^2}} 
U\left( -\frac{L_1^2 m_n^2}{4 c_1},\frac{1}{2}-c_0,\frac{c_1z^2}{L_1^2}
\right),
\\
f_n^R(z)=
N_n \frac{m_n}{2} z^{1-c_0} e^{-\frac{c_1z^2}{2L_1^2}} 
U\left( 1-\frac{L_1^2 m_n^2}{4 c_1},\frac{3}{2}-c_0,\frac{c_1z^2}{L_1^2}
\right),
\end{array}
\right\} \Rightarrow \mbox{ for } c_1 > 0, \\ 
&& \left . 
\begin{array}{l}
f_n^L(z)=
-N_n \frac{m_n}{2} z^{1+c_0} e^{\frac{c_1z^2}{2L_1^2}} 
U\left( 1+\frac{L_1^2 m_n^2}{4 c_1},\frac{3}{2}+c_0,-\frac{c_1z^2}{L_1^2}
\right),
\\
f_n^R(z)=
N_n z^{c_0} e^{\frac{c_1z^2}{2L_1^2}} 
U\left( \frac{L_1^2 m_n^2}{4 c_1},\frac{1}{2}+c_0,-\frac{c_1z^2}{L_1^2}
\right),
\end{array}
\right\} \Rightarrow \mbox{ for } c_1 < 0, 
\end{eqnarray}
where $U(a,b,z)$ is the confluent hypergeometric function and
the normalization constants $N_n$ are fixed by normalizing either
the LH or the RH profile. The linear equations of motion guarantee
that once one of the two profiles is normalized, the other one also
is. 

The masses and the possible presence of zero modes is determined
by the boundary conditions (bc). Model building in hard wall models makes
use of four different combinations of bc for fermions.
$[\pm,\pm]$ where a $+$ ($-$) means that the RH (LH) chirality has
Dirichlet bc (it vanishes) at the corresponding brane. 
The first and second signs correspond to the UV and IR
branes respectively. More complicated bc 
are sometimes needed, but they can be constructed in general from these
basic building blocks, that we would like to be able to realize in soft
wall models. The bc at the UV brane can be imposed in
exactly the same fashion as in hard wall models. On the other hand, 
the IR bc in the soft wall are fixed to the normalizability
condition and we cannot impose further bc in the
IR. As we will see, however, the choice of the sign of $c_1$ allows us
to simulate, in the soft wall, 
the effect that different IR bc have in hard wall models. 

Let us start with the analysis of \textbf{zero modes}. If we set
$m_0=0$, the two first order differential
equations decouple and we can solve for
them immediately,
\begin{equation}
f_0^{L,R} = A_{L,R} e^{\mp \int a M}
= A_{L,R} z^{\mp c_0} e^{\mp \frac{c_1z^2}{2L_1^2}}.
\end{equation}
If we choose $[+]$ UV bc, then $A_R=0$ and similarly $[-]$ implies
$A_L=0$. Thus we can only have at most one chiral zero mode. 
This chiral zero mode will be normalizable only if $c_1>0$, for a
LH zero mode, or $c_1<0$ for a RH mode.
The corresponding normalized zero mode reads,
\begin{equation}
f_0^{L,R}=\left[\frac{L_0^{1\mp 2c_0}}{2} E_{\pm c_0+\frac{1}{2}}
  \left(\pm c_1\frac{L_0^2}{L_1^2} \right) \right]^{-\frac{1}{2}}
z^{\mp c_0}
e^{\mp\frac{c_1 z^2}{2L_1^2}},
\end{equation}
where 
$E_\nu(z)= \int_1^\infty dt\, e^{-zt}/t^\nu$ is the Exponential
Integral E function.
A LH zero mode exists if $c_1>0$ and the UV bc is $[+]$,
whereas a RH zero mode exists if $c_1<0$ and the UV bc is $[-]$.
Thus, at least at the level of the zero mode content, we have the equivalence
\begin{equation}
[\pm,\pm]_{\mathrm{hard}} \Leftrightarrow 
[\pm,\mathrm{sign}(c_1)]_{\mathrm{soft}}.
\end{equation}
Once the right boundary conditions for the existence of a chiral zero
mode are imposed, we see that $c_1$ controls the exponential die-off
in the IR whereas $c_0$ controls the localization of the zero mode.
 
\begin{figure}[ht]
\includegraphics[width=.75\textwidth]{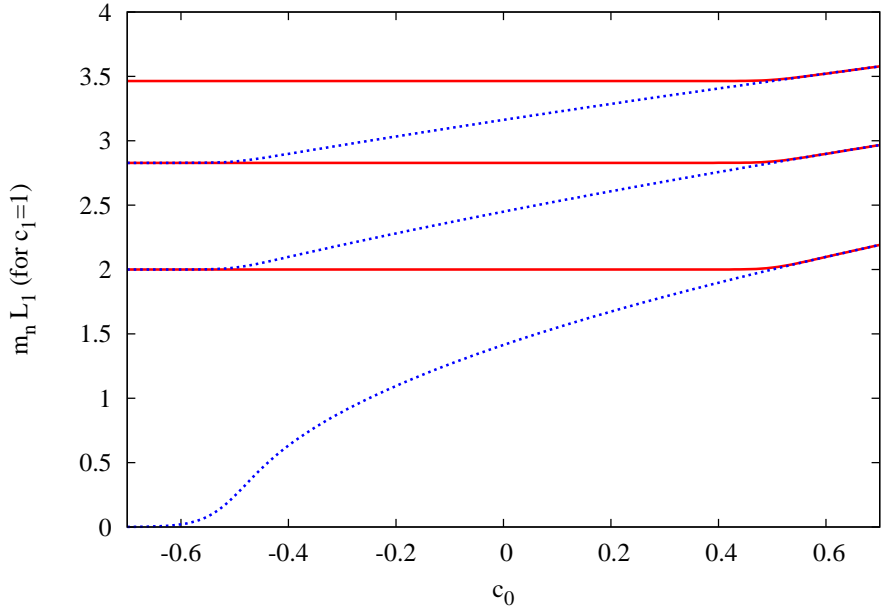}
\caption{Masses of the first three massive modes in units of $L_1^{-1}$ as
  a function of $c_0$ for $L_0/L_1 = 10^{-15}$ and $c_1=1$. The solid
  (dotted) lines correspond to $[++]$ ($[-+]$) boundary
  conditions. Opposite boundary conditions are have identical masses
  with the replacement $c_{0,1} \to - c_{0,1}$.
\label{exactmasses}}
\end{figure}

Let us now see that this identification also works at the quantitative
level for the \textbf{massive modes}.
\begin{table}[ht]
\begin{tabular}{|c|c|c|}
\hline
 & $c_0 \ll \frac{1}{2}$  & $c_0 \gg \frac{1}{2}$ \\
\hline
$[++]_{\mathrm{soft}}$ & $\frac{L_1^2m_n^2}{4 |c_1|}\approx n$ 
& $\frac{L_1^2m_n^2}{4 |c_1|}\approx n+c_0-\frac{1}{2}$ \\
$[+-]_{\mathrm{soft}}$ & $\frac{L_1^2m_n^2}{4 |c_1|}\approx
n-\frac{1}{2}-c_0$ 
& $\frac{L_1^2m_n^2}{4 |c_1|}\approx n-1$ \\
\hline \hline
 & $c_0 \ll -\frac{1}{2}$  & $c_0 \gg -\frac{1}{2}$ \\
\hline
$[--]_{\mathrm{soft}}$ & $\frac{L_1^2m_n^2}{4 |c_1|}\approx 
n-\frac{1}{2}-c_0$ & $\frac{L_1^2m_n^2}{4 |c_1|}\approx n$ \\
$[-+]_{\mathrm{soft}}$ & $\frac{L_1^2m_n^2}{4 |c_1|}\approx n-1$ &
$\frac{L_1^2m_n^2}{4 |c_1|}\approx n+c_0-\frac{1}{2}$ \\
\hline
\end{tabular}
\caption{Approximate values of the fermion KK masses in the limit
  $L_0 \ll L_1$ for different boundary conditions. Note that in the
  soft wall, the second sign in the boundary condition gives the sign
  of $c_1$ and that $n=1,2,\ldots$. The lightest modes ($n=1$)
for $[+-]$ and $[-+]$ boundary conditions require subleading terms
that have been neglected in this table, see text for details.  
 \label{approximate:fermion:masses}}
\end{table}
Using the small $z$ limit of the confluent hypergeometric functions,
Eqs. (13.5.6-13.5.12) of \cite{Abramowitz}, 
\begin{equation}
U(a,b,z) \sim \left \{ \begin{array}{l}
\frac{\Gamma(b-1)}{\Gamma(a)}z^{1-b}, \quad b>1, \\
\frac{\Gamma(1-b)}{\Gamma(1+a-b)}, \quad b<1,
\end{array}\right.
\end{equation}
and assuming $L_0/L_1 \ll 1$, we obtain the approximate expressions
for the masses of the KK modes for different values of the UV bc and signs
of $c_1$ as shown in Table~\ref{approximate:fermion:masses}.
In the case of $[+-]$ and $[-+]$ boundary conditions subleading terms,
which are particularly important for the lightest mode, have been
neglected in the table. 
Including those terms, we obtain the following approximate
expressions for the masses
\begin{equation}
m_1^2 \approx \frac{4 |c_1|}{L_1^2}
\frac{|c_1|^{c_0-\frac{1}{2}}}{\Gamma(c_0-1/2)} \left(
  \frac{L_0}{L_1} \right)^{2c_0 -1},
\quad \mbox{ for }[+-], ~c_0 >1/2+\epsilon,
\end{equation}   
and
\begin{equation}
m_1^2 \approx \frac{4 c_1}{L_1^2}
\frac{c_1^{-c_0-\frac{1}{2}}}{\Gamma(-c_0-1/2)} \left(
  \frac{L_0}{L_1} \right)^{-2c_0 -1},
\quad \mbox{ for }[-+], ~c_0 <-1/2-\epsilon,
\end{equation}   
with $\epsilon \approx 0.1$. These are 
ultralight modes for $[+-]$, $c_0>1/2$ and $[-+]$,
$c_0<-1/2$ similar to the ones that appear in hard wall models with
the corresponding twisted boundary
conditions~\cite{DelAguila:2001pu}. The
accuracy of these approximations can be checked in
Fig.~\ref{exactmasses}, where we show the exact masses for the
different boundary conditions. The asymptotic limits, the scaling
$m_n\sim \sqrt{n}$ and the presence of ultralight modes for twisted
boundary conditions is apparent from the figure.  

\begin{figure}[ht]
\includegraphics[width=.75\textwidth]{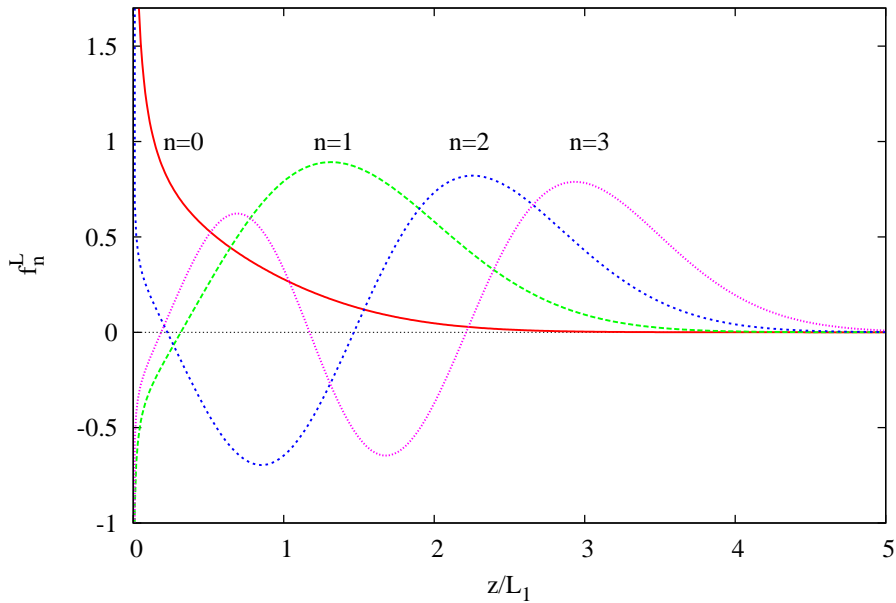}
\caption{Left handed profiles of the zero mode and first three massive
  KK modes of a bulk fermion with $[++]$ boundary conditions,
  $c_0=0.4$ and $c_1=1$. Heavier KK modes propagate deeper in the IR.
\label{fnLpp:plot}}
\end{figure}

The $z$-dependent Dirac mass shields the fermions from the deep IR. Of
course, the heavier a particular KK mode is, the less it is shielded
from the IR. This can be seen in Fig.~\ref{fnLpp:plot}, where we show
the profiles of a LH zero mode and its first three massive KK
modes for $[++]$ boundary conditions,
with heavier modes propagating deeper in the IR. This has
important phenomenological 
consequences. It has been shown that fields propagating
sufficiently deep in the IR become eventually
strongly coupled~\cite{Karch:2006pv,Falkowski:2008fz}. 
This strong coupling
does not affect the Standard Model (SM) fields, 
as they do not propagate deep enough in
the IR (not even the top), but will signal the loss of perturbativity
for heavy enough KK modes. In fact, we will see in the next section
that, due to this effect,
the contribution of fermionic KK modes to some EW
observables decouples more slowly than one might naively expect.

The KK expansion of bulk fermions with a bulk Dirac mass of the form
 in Eq.~(\ref{bulk:mass}) makes perfect sense independently
of the Yukawa couplings to the bulk Higgs. Our goal in the rest of this
article is to use this expansion to analyze the phenomenological
implications of bulk fermions in soft wall models. We will do that by
treating EWSB
perturbatively, an
approximation that will be carefully checked. This approach
has the advantage that we can now trivially implement all the
complications involved in fermion masses, from the generation of a
large top mass to
the generation of light fermion masses and their implication in flavor
constraints of the model
without having to resort to common bulk Dirac masses. 

\subsection{Fermion couplings}

\begin{figure}[ht]
\includegraphics[width=.65\textwidth]{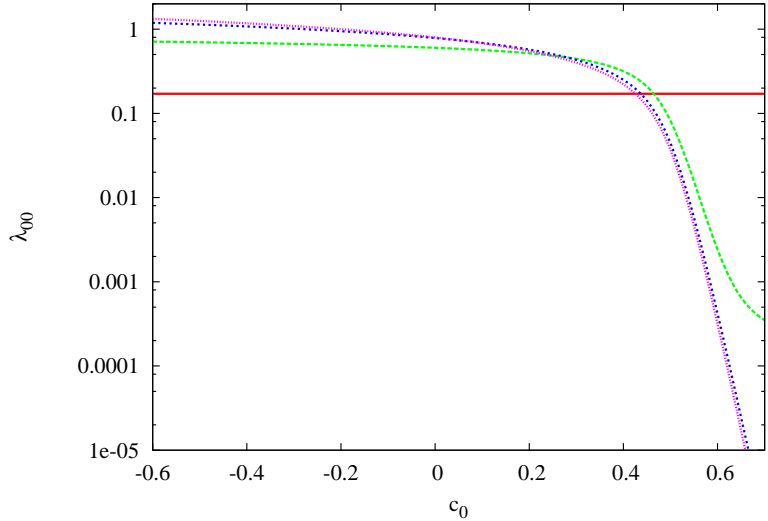}
\caption{
Effective four-dimensional Yukawa coupling for fermion zero modes with
$\lambda_5=\sqrt{L_0}$ and $c_1^L=-c_1^R=1$ as a function of
$c_0^L=-c_0^R\equiv c_0$ for different Higgs profiles. The different
curves correspond, from bottom to top on the left hand side, 
to $\alpha=1,~1.33,~1.66,~2$.
For $\alpha \gtrsim 1.5$ the zero mode spectrum is almost insensitive
to the exact value of the Higgs profile.
\label{lambda4:plot}}
\end{figure}

The KK expansion we have performed in the previous section allows us
to compute the couplings we will need to investigate the EW
and flavor implications of bulk fermions in the soft wall. One of the
most relevant couplings
is the Yukawa coupling
between two fermions and a scalar.
In the approximation we are considering, these couplings 
give the main contribution to 
SM fermion masses and mixings and also fix the mixings among fermion
KK modes. These mixings in turn determine the fermion 
effects on EW precision observables and on flavor violating processes. 
Let us consider two bulk fermions $Q(x,z)$ and $T(x,z)$ coupled to a
bulk scalar $\phi(x,z)$. We assume the scalar acquires a $z$ dependent
vev $\langle \phi \rangle = f_\phi(z) v$, with $v$ a constant with
dimension of mass and $f_\phi(z)$ normalized as
\begin{equation}
1=\int_{L_0}^\infty dz\, a^3 e^{-\Phi} f_\phi^2(z).
\end{equation}
When we identify the scalar with the Higgs responsible for EWSB, 
this normalization ensures that $v=174$ GeV, up to
corrections of order $v^2 L_1^2$.
The part of the action involving the coupling between these three
fields reads
\begin{eqnarray}
S_\mathrm{Yuk}&=& 
\int d^5x\, 
\sqrt{g} e^{-\Phi} \Big[ \lambda_5 \bar{Q} \phi T +
  \mathrm{h.c.}
\Big] 
=
\int d^5x\, 
a \Big[ \lambda_5 \bar{q} \phi t +
  \mathrm{h.c.}
\Big] 
=\int d^5x\, 
a \Big[ \lambda_5 v f_\phi  \bar{q}   t +
  \mathrm{h.c.}
\Big] 
\nonumber \\
&=&
\int d^4x\, v 
\sum_{mn}\Bigg \{ \lambda_5 
\left[\int dz  a f_\phi f_m^{qL} f_n^{tR} 
 \right]
\bar{q}^{(m)}_{L} t_R^{(n)} +
(L \leftrightarrow R)
+  \mathrm{h.c.}
\Bigg\} 
\nonumber \\
&=&
\int d^4x\, v 
\sum_{mn}\Bigg \{ 
\lambda_{mn}^{qt} \bar{q}^{(m)}_{L} t_R^{(n)} 
+
\lambda_{mn}^{tq} \bar{t}^{(m)}_{L} q_{R}^{(n)} 
+  \mathrm{h.c.}
\Bigg\}, \label{y:coupling}
\end{eqnarray}
where $\lambda_5$ is a five-dimensional Yukawa coupling with mass
dimensions $[\lambda_5]=-1/2$, naturally expected to be of order
$\lambda_5 \sim \sqrt{L_0}$. In the last equality 
we have defined the effective (dimensionless)
four-dimensional Yukawa couplings between the different fermion KK
modes. These are given by the five-dimensional Yukawa coupling times
the overlap of the corresponding fermion KK modes with the scalar profile.
Following the procedure in the KK expansion of fermions, we have defined
$q(x,z)\equiv a^{2} e^{-\Phi/2} Q(x,z)$ 
and $t(x,z)\equiv a^{2} e^{-\Phi/2} T(x,z)$.
Assuming a power-law scalar profile (see Appendix~\ref{bosons})
\begin{equation}
f_\phi(z) = \frac{L_1}{L_0^{\frac{3}{2}}} \left[
  \frac{2}{\Gamma(\alpha-1,L_0^2/L_1^2)}\right]^{\frac{1}{2}} \left(
  \frac{z}{L_1} \right)^\alpha, \label{f:phi}
\end{equation}
and identical localization for the LH and RH fermion zero modes
($c_{0,1}^L = -c_{0,1}^R$), we show the effective zero mode Yukawa
coupling for different values of the Higgs profile in
Fig.~\ref{lambda4:plot}. The result is
sensitive to the Higgs profile for $\alpha \lesssim 1.5$ but becomes
essentially insensitive and very similar at the qualitative and
quantitative level to the result in the hard wall for $\alpha \gtrsim
1.5$. 

The another very important 
coupling for the phenomenological implications of
bulk fermions is the coupling of
fermion  and gauge boson KK modes. It comes from the gauge
covariant derivative in the kinetic term of fermions and can be
written as
\begin{eqnarray}
S_{A\psi \psi}&=& 
\int d^5x \, g_5 \bar{\psi} \cancel{A} \psi 
= 
\int d^4 x \, \sum_{mnr} g_5 \left[ \int dz f_m^L f_n^L f_r^A
  \right] \bar{\psi}^{(m)}_L \cancel{A}_r \psi^{(n)}_L + (L \to R) \nonumber \\
&=& 
\int d^4 x \, \sum_{mnr} g_4^{{m_L}{n_L}r} 
\bar{\psi}^{(m)}_L \cancel{A}_r \psi^{(n)}_L + (L \to R), \label{fA:coupling}
\end{eqnarray}
where $g_5$ is the five-dimensional coupling constant, with mass
dimension $[g_5]=-1/2$ and in the last equality we have defined the effective
four-dimensional coupling between the fermionic modes $\psi_L^{(m)}$,
$\psi_L^{(n)}$ and the $r$-th gauge boson KK mode.
Of  particular relevance is the coupling of fermion zero modes to
gauge boson KK modes,
\begin{equation}
g_n(c_0,c_1) \equiv g_4^{{0}_L{0}_Ln}= g_5 \int dz\, (f_0^L)^2 f_n^A,
\end{equation}
which we show, in units of $g_0$ for the first three gauge boson KK
modes, as a function of $c_0$ in Fig.~\ref{gnbyg0:plot} for $c_1=1$. 
The KK expansion of gauge bosons in our soft wall background is
discussed in Appendix~\ref{bosons}. It is related to the fermionic
expansion with the identification
\begin{equation}
f^L_n(c_0=1/2,c_1=1,z)= \sqrt{a} e^{-\Phi/2} f_n^A.
\end{equation}
In particular we have
\begin{equation}
g_n(c_0=1/2,c_1=1) = g_5  f_0^A \int dz\, a e^{-\Phi} f_0^A f_n^A =
g_0 \delta_{n0},
\end{equation}
where we have used the fact the $f_0^A$ is $z$ independent,
$g_5 f_0^A = g_0$ and orthonormality of the gauge
boson KK modes, see Eq.~(\ref{orthonormality:gauge}). 
Thus, for $c_0=1/2$ and $c_1=1$ the couplings to all the
gauge boson KK modes vanish. Similarly, for $c_0 \gtrsim 1/2$, the
fermion zero modes are effectively localized towards the UV brane and their
coupling to the gauge boson KK modes becomes independent of the exact
localization (the value of $c_{0,1}$). For instance, the coupling to
the first gauge boson KK mode becomes
\begin{equation}
g_1(c_0 > \frac{1}{2}+\epsilon) \approx -\frac{g_0}{\sqrt{2\log
    \frac{L_1}{L_0}}} \approx - 0.12 g_0, \quad \mbox{ (for $L_0/L_1 \sim
  10^{-15}$)}. 
\end{equation}
\begin{figure}[ht]
\includegraphics[width=.65\textwidth]{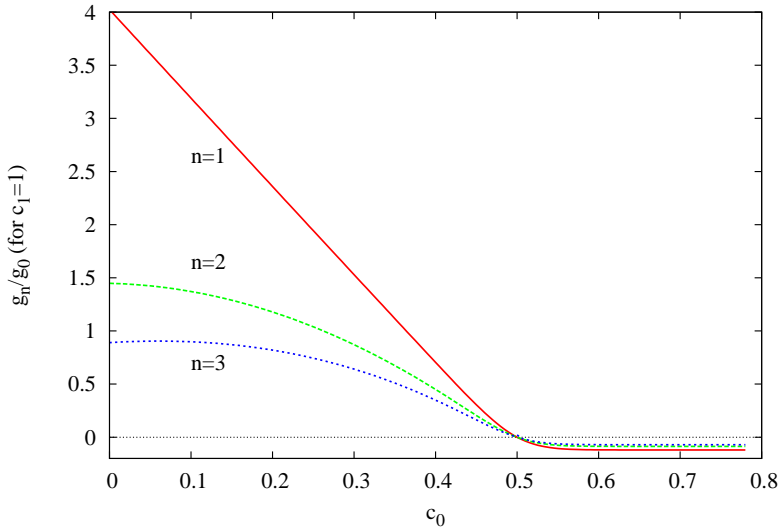}
\caption{
Coupling of a LH fermion zero mode to the first three gauge boson KK
modes in units of the coupling to the gauge boson zero mode, as a
function of $c_0$ for fixed $c_1=1$.
\label{gnbyg0:plot}}
\end{figure}
This universality of couplings for fermions localized towards the UV
brane is the basis of the success of flavor physics in models with
warped extra dimensions with a hard wall.
The so called RS GIM mechanism in hard wall models stands for the fact
that FCNC processes involving light quarks 
are suppressed by either light quark masses or by
small CKM mixing angles. The origin of such suppression comes from the
fact that the deviation from universality of the couplings to the
gauge boson KK modes scale like the Yukawa couplings
\begin{figure}[ht]
\includegraphics[width=.65\textwidth]{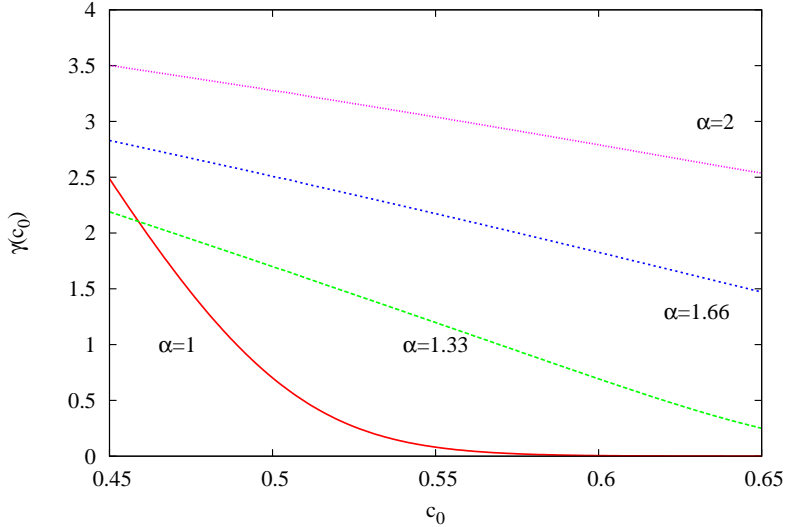}
\caption{
$\gamma(c_0)^\mathrm{soft}$ as defined in Eq.~(\ref{gamma:soft}).
This quantity measures the scaling of the non-universality of
couplings of light fermions to gauge boson KK modes with the masses of
the light fermions.
We have fixed $c_1^q=-c_1^t=c_1$, $c_0^q=c_0^t=c_0$,
$\lambda_5=\sqrt{L_0}$ and shown different lines for different
Higgs profiles.
\label{g1scaling:plot}}
\end{figure}
\begin{equation}
\frac{g^{L,R}_n}{g_0} \approx \mathrm{const.} + f_{c_{L,R}}^2 \gamma(c_{L,R}),
\end{equation}
where $c_{L,R}$ is the localization parameter of the corresponding LH
or RH fermion zero mode, $\gamma(c)$ is a slowly varying function of
$c$, expected to be of order $\sqrt{\log L_1/L_0}\approx 5-6$ in the
hard wall and $f_{c_{L,R}}$ determine the effective Yukawa couplings
as
\begin{equation}
\lambda_{00} \sim f_{c_L} f_{c_R}.
\end{equation}
To test how close to this property we are in our soft wall models, we
 computed the following quantity
\begin{equation}
\gamma(c_0)^{\mathrm{soft}} \equiv \frac{[g_n(c_0,c_1) - g_n(c_0\to
  \infty,c_1)]/g_0}{
  \lambda_{00}^{qt}(c_0^q=-c_0^t=c_0,c_1^q=-c_1^t=c_1)},
\label{gamma:soft}
\end{equation}
which is the equivalent  of $\gamma(c)$ in the hard wall 
for identical localization of left and right
components. The result for the
coupling to the first gauge boson KK mode, for $c_1=1$
and $\lambda_5=\sqrt{L_0}$ is displayed in Fig.~\ref{g1scaling:plot},
for different values of the Higgs profile. It is obvious that for
$\alpha \gtrsim 1.5$, for which we can obtain hierarchical fermion
masses through wave function localization, similar scaling with the
masses as the one that leads to a RS GIM protection occurs in the soft
wall. Thus, one would expect that models with a soft wall behave
 in a very similar way as hard wall models
do regarding flavor physics~\cite{flavor,Csaki:2009bb}.

\subsection{Validity of the approximation}

The validity of the approximation we are employing in this work,
namely a perturbative treatment of EWSB, has been recently subject to
debate in
hard wall models (see for
instance~\cite{Goertz:2008vr,Csaki:2009bb}). In such models, it is
clear that, if one was able to include the full tower of KK modes and
diagonalized the infinite resulting mass matrix, the results should be
equivalent to including EWSB effects exactly
through the equations of motion. The rapid increase in KK masses also
guarantees that in general, except in certain cases in which the induced
off-diagonal masses are very large, the first few modes give a good
enough approximation for the required level of precision.   
A perturbative treatment of EWSB is not useful only for soft wall
models. In fact, even in hard wall models, there is only a few special
cases, notably a brane localized Higgs and models of gauge-Higgs
unification, in which EWSB effects can be included analytically.
The validity of a perturbative treatment of EWSB 
is not so obvious in soft wall models, since the extra dimension
extends to infinity and Higgs effects in the deep IR could modify the
equations of motion in a way that cannot be reproduced with a finite
number of modes computed in the absence of EWSB 
and a perturbative inclusion of the latter. This is in fact
what happens in the case that the fermion Dirac mass is constant in
the extra dimension so that the KK expansion in the absence of
EWSB lacks a zero mode that should make up for most of the lightest
mode in the full solution. With our assumption of fermion Dirac
masses, however, the KK expansion in the absence of EWSB is well
defined and all fields, including the zero modes are shielded from the
deep IR. We can therefore expect, in general, EWSB effects to be a small
perturbation and a finite number of KK modes to give a good enough
approximation. Nevertheless, it is important to check the validity of
the approximation at the quantitative level, especially in the case of
the top quark, which is expected to suffer the strongest
deviations. Even more so in soft wall models due to the fact we
mentioned above that heavier KK modes propagate deeper in the IR
and therefore get larger mass mixings.

We have checked the validity of the approximation both analytically
and numerically. First, we have considered the case that can be solved
analytically of common \textit{constant} bulk Dirac masses and a
quadratic Higgs profile. Let us consider again 
two bulk fermions $Q(x,z)$ and $T(x,z)$ but let us include the effect
of the Yukawa coupling, Eq.~(\ref{y:coupling}) in the KK expansion.
The equations of motion derived from the corresponding action read
\begin{equation}
i\cancel{\partial} 
\begin{pmatrix} q_{R,L} \\ t_{R,L} \end{pmatrix}
\mp \partial_5
\begin{pmatrix} q_{L,R} \\ t_{L,R} \end{pmatrix}
- a \begin{pmatrix} M_q(z) & m(z) \\ m(z) & M_t(z) \end{pmatrix} 
\begin{pmatrix} q_{L,R} \\ t_{L,R} \end{pmatrix}
=0,
\end{equation}
where $M_{q,t}$ are our constant plus $z$ dependent mass for the bulk
fermions (with parameters $c^{q,t}_{0,1}$) and $m(z)\equiv \lambda_5 v
f_\phi(z)$. Assuming a quadratic profile for the scalar and
$c_0^q=c_0^t\equiv c_0$, the mass matrix reads
\begin{equation}
 a \begin{pmatrix} M_q(z) & m(z) \\ m(z) & M_t(z) \end{pmatrix} 
= \frac{c_0}{z} \begin{pmatrix} 1 & 0 \\ 0 & 1 \end{pmatrix}
+ \frac{z}{L_1^2} \begin{pmatrix} c_1^q & c_m \\ c_m &
  c_1^t \end{pmatrix}.
\end{equation}
Let us now define the $2\times 2 $ unitary matrix that diagonalizes
the second term
\begin{equation}
U^\dagger \begin{pmatrix} c_1^q & c_m \\ c_m & c_1^t \end{pmatrix} U
= \begin{pmatrix} c_1^+ & 0 \\ 0 & c_1^- \end{pmatrix}.
\end{equation}
Then the equations of motion for the rotated fields
\begin{equation}
\begin{pmatrix} \psi^+ \\ \psi^- \end{pmatrix}
\equiv U^\dagger \begin{pmatrix} q \\ t \end{pmatrix},
\end{equation}
are decoupled
\begin{equation}
i \cancel{\partial} \psi_{L,R}^i \pm \partial_5 \psi_{R,L}^i - \left(
  \frac{c_0}{z} +\frac{c_1^i z}{L_1^2} \right)\psi_{R,L}^i = 0, \quad i=+,-,
\end{equation}
which can be solved analytically. One has to be careful that the
boundary conditions mix both fields and therefore the physical modes
live in the expansion of both of them
\begin{equation}
\psi_{L,R}^i = \sum_n f_n^{i\, L,R} \psi^{(n)}_{L,R},
\end{equation}
where $\psi^{(n)}$ is independent of $i$. The orthonormality condition
is then
\begin{equation}
\int_{L_0}^\infty dz\, \left[ f_n^{+\,L}f_m^{+\,L}+
  f_n^{-\,L}f_m^{-\,L} \right]
=\int_{L_0}^\infty dz\, \left[ f_n^{+\,R}f_m^{+\,R}+
  f_n^{-\,R}f_m^{-\,R} \right]=\delta_{nm}.
\end{equation}
We have compared the exact fermion masses and the couplings to gauge
bosons with the ones we obtain using our perturbative treatment of
EWSB. We have compared them for different values of $c_0$ and
$c_1^{q,t}$, with $\lambda_5$ (equivalently $c_m$) 
fixed to reproduce the top mass for the
lightest mode. The result of such comparison is that both the masses
and the couplings agree to better than per mille level for values of
$\lambda_5$ below the strong coupling limit, with the inclusion of
just a few KK modes. Only when $\lambda_5$ is
so large that the theory becomes non-perturbative close to the scale
of the mass of the first KK mode the departure gets close to $\sim
10\%$.

We have been able to test the accuracy of our approximation in the
case of common \textit{constant} bulk Dirac masses. However, this is
not the most general situation we will deal with in the study of the
phenomenological implications of bulk fermions. We have also checked
the case of different Dirac masses by numerically solving the set of
coupled differential equations. Again the result for both masses and
couplings is in excellent agreement with our approximation for
perturbative values of $\lambda_5$.

\section{Electroweak Constraints on Soft Wall
  Models \label{ew:constraints}} 

As an application of our formalism, we investigate in this section
the EW constraints on soft wall models. We consider a minimal realistic
set-up with a custodially symmetric 
$SU(2)_L\times SU(2)_R\times U(1)_X$ bulk gauge symmetry broken to the
SM gauge group $SU(2)_L\times U(1)_Y$ on the UV brane.

The bosonic action is given by
\begin{eqnarray}
\label{bosonicL}
S_b&=&
\int d^5x\,\sqrt{g}\,\mathrm{e}^{-\Phi}
\Bigg[ -\frac{1}{4}
  L_{MN}^a\,L^{a\,MN} - \frac{1}{4}  R_{MN}^a\,R^{a\,MN} - \frac{1}{4}
  X_{MN} X^{MN} \nonumber \\ 
& &\phantom{\int d^5x\,\sqrt{g}\,\mathrm{e}^{-\Phi}} 
+ \mathrm{Tr}\,[(D_M H)^\dagger D^M H] - V(H)\Bigg]
-\int d^4x\,\sqrt{-g_{UV}}{\rm e}^{-\Phi}\,V_{UV}(H)\,, 
\end{eqnarray}  
where $a=1,2,3$, $L^a_M$, $R_M^a$ and $X_M$ are the $SU(2)_L$, $SU(2)_R$ and
$U(1)_X$ gauge fields, respectively, 
and $H$ is the bulk Higgs boson that transform as 
an $SU(2)_L\times SU(2)_R$ bidoublet
\begin{equation}
H(x,z)=\frac{1}{\sqrt{2}}\left(
\begin{array}{cc}
\phi_0^*(x,z) & \phi^+(x,z) \\
-\phi^-(x,z) & \phi_0 (x,z) 
\end{array}
\right)\,.
\end{equation}
$V(H)$ and $V_{UV}(H)$ are the bulk and UV brane Higgs
potentials. Choosing appropriately these potentials we can obtain a
power-like profile for the Higgs vev (see Appendix~\ref{bosons})
\begin{equation}
\langle H \rangle =
\frac{f_\phi(z)}{\sqrt{2}}
\begin{pmatrix} v & 0 \\ 0 & v \end{pmatrix},
\end{equation}
where $f_\phi(z)$ has been defined in Eq.~(\ref{f:phi}) and, as we
show below,  
$v=174$ GeV up to corrections of order $v^2 L_1^2$.
The UV boundary conditions satisfied by the gauge
fields are 
\begin{equation}
\partial_5 L^a_{\mu} \big|_{L_0} =
\partial_5 B_{\mu} \big|_{L_0} =
R_{\mu}^{1,2} \big|_{L_0} =
Z_{\mu}^{\prime}\big|_{L_0} =0\,,
\end{equation}
where we have defined 
\begin{equation}
B_{\mu}=\frac{g_XR_{\mu}^3+g_5 X_{\mu}}{\sqrt{g_5^2+g_X^2}},
\quad
Z_{\mu}^{\prime}=\frac{g_5 R_{\mu}^3-g_X X_{\mu}}{\sqrt{g_5^2+g_X^2}},
\end{equation}
with $B_\mu$ the hypercharge gauge boson, $g_5$ the gauge coupling of
$SU(2)_{L}$ and $SU(2)_R$ which are taken equal to maintain the
the $L\leftrightarrow R$ symmetry that protects the $Z b \bar{b}$
coupling~\cite{Agashe:2006at} and $g_X$ the $U(1)_X$ gauge
coupling. The gauge couplings associated to $B_\mu$ and
$Z_\mu^\prime$ are, respectively
\begin{equation}
g_5^\prime = \frac{g_5 g_X}{\sqrt{g_5^2+ g_X^2}},
\quad
g_{Z^\prime} = \sqrt{g_5^2+ g_X^2},
\end{equation}
whereas the associated charges are
\begin{equation}
Y=T^3_R+Q_X,\quad
Q_{Z^\prime} = \frac{g_5^2 T^3_R - g_X^2 Q_X}{g_5^2 + g_X^2}.
\end{equation}

\subsection{Bosonic contribution to electroweak precision observables}

In this section we review the bosonic contribution to EW
precision observables in soft wall
models~\cite{Falkowski:2008fz,Batell:2008me}. 
If the light SM fermions are localized towards the UV brane, models
with warped extra dimension, including soft wall models, fall in the
class of universal new physics models~\cite{Barbieri:2004qk}. In that
case, all relevant constraints can be obtained in terms of four
oblique parameters, which are computed from the quadratic terms in the
effective Lagrangian of the interpolating fields that couple
universally to the SM fermions. The holographic
method~\cite{Barbieri:2003pr} gives the most straight-forward
calculation of the oblique parameters as it gives directly the
effective Lagrangian for the interpolating fields. In the spirit of
the approximations we have used with fermions, we will include EWSB
effects perturbatively, which allows us to consider arbitrary Higgs
profiles. The leading order in the expansion in the EWSB scale $v$
does not give the correction to the $T$ parameter (which requires
$v^4$ terms) but we nevertheless know that the tree level contribution to
the $T$ parameter vanishes due to the custodial symmetry.\footnote{This can
be checked for particular Higgs profile~\cite{Batell:2008me}.} The
leading correction to all other three oblique parameters can be
safely computed perturbatively.

The procedure can be performed as follows. First we set to zero the
fields that are not sourced on the UV brane, leaving only the SM gauge
bosons. The relevant part of the action reads
\begin{equation}
S=\int d^5x \sqrt{g}\mathrm{e}^{-\Phi}\left\{  -\frac{1}{4g_5^2}
L_{MN}^aL^{a\,MN} - \frac{1}{4g_5^{\prime\, 2}}  B_{MN}B^{MN} 
+ \frac{v^2 f_\phi^2}{4} \Big[ (L^b_M)^2 + (L^3_M-
B_M)^2\Big] \right\}, 
\end{equation}
where in this section, we are using non-canonically normalized fields.
The term in square brackets is the mass term due to EWSB, that we will
treat as a perturbation.~\footnote{If we were to include the EWSB
  effects exactly, it would be advantageous to go the the vector and
  axial basis, $V,A=(L\pm R)/\sqrt{2}$, but this is not necessary if
  EWSB is treated as a perturbation.} 
Going to 4D momentum space, and considering the $\mu$ components of
the gauge fields, we define
\begin{equation}
L^a_\mu(p,z) = f(p,z) \bar{L}^a_\mu(p),
\quad
B_\mu(p,z) = f(p,z) \bar{B}_\mu(p),
\end{equation}
with $f(p,z)$ fixed by the bulk equations of motion, which are
identical for $L^a$ and $B$. With the boundary
condition $f(p,L_0)=1$,
\begin{equation}
f(p,z)=
U\left( - \frac{p^2 L_1^2}{4},0,\frac{z^2}{L_1^2} \right) 
\Big/
U\left( - \frac{p^2 L_1^2}{4},0,\frac{L_0^2}{L_1^2} \right). 
\end{equation}
Integrating out the bulk we obtain an effective action for the 4D
interpolating fields 
\begin{equation}
S_{\mathrm{hol.}}=
-\frac{1}{2} \int \frac{d^4p}{(2\pi)^4} \eta^{\mu\nu}
\left\{ 
\bar{L}^b_\mu \Pi_{+-}(p^2) \bar{L}^b_\nu 
+ \bar{L}^3_\mu \Pi_{33}(p^2) \bar{L}^3_\nu
+ \bar{B}_\mu \Pi_{BB}(p^2) \bar{B}_\nu 
+ 2 \bar{L}^3_\mu \Pi_{3B} \bar{B}_\nu
\right\} + \ldots
\end{equation}
where the different form factors read
\begin{eqnarray}
\Pi_{+-}(p^2) &=& 
\Pi_{33}(p^2) =\frac{1}{g_5^2}\Pi_0(p^2) - \Pi_\phi(p^2),
\label{Pi:aa}\\
\Pi_{BB}(p^2)
&=& \frac{1}{g_5^{\prime\,2}}\Pi_0(p^2) - \Pi_\phi(p^2),
\label{Pi:BB}\\
\Pi_{3B}(p^2) &=& \Pi_\phi(p^2)\,,
\label{Pi:3B}
\end{eqnarray} 
with the EW symmetry preserving term
\begin{equation}
\Pi_0(p^2) \equiv
 e^{-\Phi} \partial_5 f(p,z) \Big|_{z={L_0}}
\approx 
p^2 L_0 \left( \log \frac{L_1}{L_0} - \frac{\gamma_E}{2} \right)
+ \frac{p^4}{2} L_0 \frac{\pi^2 L_1^2}{24} + \ldots ,
\end{equation}
where $\gamma_E\approx 0.577$ is the Euler-Mascheroni constant. 
In the second expression we have expanded in powers of $p^2$ and
assumed $L_0/L_1 \ll 1$. The EWSB term is
\begin{equation}
\Pi_\phi(p^2) = \frac{v^2}{2}
\int dz\, a^3 e^{-\Phi} f_\phi^2 f^2(p,z) = \frac{v^2}{2} +
\mathcal{O}(p^2). 
\end{equation} 
The gauge couplings and the EWSB scale are fixed by the following
conditions
\begin{equation}
\Pi_{+-}^\prime (0)=\frac{1}{g^2},\quad
\Pi_{BB}^\prime(0) =\frac{1}{g^{\prime\,2}}, \quad
\Pi_{+-}(0)= - \frac{ (174~\mathrm{GeV})^2}{2},
\end{equation}
where a prime here denotes derivative with respect to $p^2$. They 
imply, up to corrections $\mathcal{O}(v^2 L_1^2)$,
\begin{equation} 
\frac{g_5}{g}= \frac{g_5^\prime}{g^\prime}= \sqrt{
L_0 \left( \log \frac{L_1}{L_0} -
    \frac{\gamma_E}{2} \right)}, 
\end{equation}
and
\begin{equation}
v =  174 ~\mathrm{GeV}. 
\end{equation}
The oblique parameters are defined in terms of these form
factors. Recall that the
first equality in Eq.~(\ref{Pi:aa}) does not give the leading
correction to the $T$ parameter, which appears at order $v^4$ in the
form factors. The other three relevant oblique parameters are defined
by 
\begin{eqnarray}
S &=& 16 \pi \Pi^\prime_{3B}(0)=-16 \pi g^2 \Pi_\phi^\prime(0),
\label{Sparameter} \\
W &=& \frac{g^2 m_W^2}{2} \Pi_{33}^{\prime\prime}(0), \\
Y &=& \frac{g^{\prime\,2} m_W^2}{2} \Pi_{BB}^{\prime\prime}(0).
\end{eqnarray}
Inserting the corresponding coefficients we obtain, for $W$ and $Y$,
\begin{equation}
W=Y=\frac{g^2 \pi^2}{96\left(\log \frac{L_1}{L_0} -
  \frac{\gamma_E}{2}\right)} (v L_1)^2 + \ldots
\end{equation}
which is volume ($\log (L_1/L_0)-\gamma_E/2$)
suppressed and therefore leads to a very mild
constraint on $L_1$. 
\begin{figure}[ht]
\includegraphics[width=.65\textwidth]{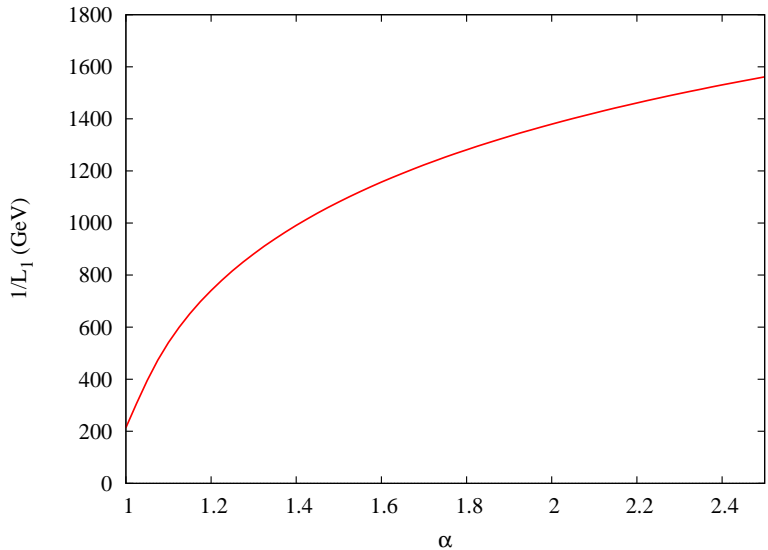}
\caption{$L_1^{-1}$ as a function of the Higgs localization parameter 
$\alpha$ such that $S=0.2$ using Eq.~(\ref{Sparameter}).
\label{s}}
\end{figure}
The $S$ parameter on the other hand can in
general only be computed numerically, except for particular values of
the Higgs localization parameter $\alpha$. In general it is not volume
suppressed so it gives a stronger constraint that $W$ and
$Y$. This is easy to understand in an alternative derivation in which
one integrates out the physical heavy KK modes to obtain an effective
Lagrangian that is not oblique but has corrections to the gauge boson
masses, which are volume enhanced, to vertex fermion-gauge couplings,
 which are order one, and to four-fermion interactions, which are
volume suppressed. If one then introduces field redefinitions so that
an oblique Lagrangian is obtained, all three types of corrections
affect the $T$ parameter (which would be the leading constraint if it
were not for the custodial symmetry that makes the total contribution
to cancel). Only vertex corrections and four-fermion interactions
enter in the $S$ parameter, which is therefore not expected to have
any volume enhancement or suppression. Finally only the four-fermion
interactions enter $W$ and $Y$, which explains why they are volume
suppressed and therefore less constraining in general. We have
computed the $S$ parameter as a function of the Higgs localization
parameter $\alpha$. In Fig.~\ref{s} we show, as an example, the value
of $L_1^{-1}$ that will result in a value $S=0.2$ as a function of
$\alpha$. We have checked that our numerical result agrees exactly
with the analytic results given in~\cite{Batell:2008me} for the cases
$\alpha=1$ and $\alpha=2$ to leading order in $v^2 L_1^2$. For
$\alpha=1$,  the Higgs profile times the metric is flat in the extra
dimension and no vertex corrections are generated. Thus, the $S$
parameter only receives the volume suppressed contribution from
four-fermion interactions, which results in a very mild constraint.

\subsection{Fermionic contributions to electroweak observables}

Our calculation in the previous section showed that the bosonic
sector is less constrained by EW precision tests in soft wall models
than in hard wall models. The large top mass however makes it
reasonable that fermionic contributions to EW precision observables,
most notably the $T$ parameter and the $Z b \bar{b}$ coupling,
which have been 
neglected so far can be relevant. In fact, they are needed for light
KK modes to be allowed, as the zero value of the tree level $T$
parameter is not compatible with a relatively large value of $S$. 

We consider a minimal fermionic content compatible with the
custodial and $LR$ symmetry that protect the $T$ parameter and $Z b
\bar{b}$ coupling, respectively. Ignoring the bottom or lighter quark
masses, the relevant
fermionic sector consists of an $SU(2)_L\times SU(2)_R$ bidoublet
$\psi_{(2,2)}=(X,Q)$ and an $SU(2)_L\times SU(2)_R$ singlet $T$, both
with $Q_X=2/3$,
\begin{equation}
\psi_{(2,2)}=\left(
\begin{array}{cc}
X^u[-+] & Q^u[++] \\
X^d[-+] & Q^d[++] 
\end{array} 
\right)_{\frac{2}{3}},\quad  T[--]_{2/3}\,,
\end{equation}
where the subscript denotes the $U(1)_X$ charge and we have written
explicitly the boundary conditions in soft-wall notation so that the
second sign corresponds to the sign of the corresponding $c_1$. 
From the SM point of view
$X$ and  $Q$ are $SU(2)_L$ doublets with 
hypercharges $7/6$ and $1/6$ and they have $T^3_R=
1/2$ and $-1/2$, respectively. 
The bc are chosen such that $Q$ has a left handed and $T$
has a right handed zero mode which correspond to the SM top sector
$q_L$ and $t_R$. $X$ have no zero modes but have the IR bc fixed by
the bulk gauge symmetry.

The couplings of the fermions to the gauge fields and the Yukawa
couplings are given in Eq.~(\ref{fA:coupling}) and
Eq.~(\ref{y:coupling}). The latter are given in terms of the fermion
profiles by
\begin{equation}
S_{\mathrm{Yuk}} = \int d^4x\, v \sum_{mn} \left \{
\big[
\lambda^{qt}_{mn} \bar{q}^{u\,(m)}_L + \lambda^{xt}_{mn} \bar{x}^{d\,(m)}_L 
\big] t^{(n)}_R +
\bar{t}^{(m)}_L 
\big[
\lambda^{tq}_{mn} q^{u\,(n)}_R + \lambda^{tx}_{mn} x^{d\,(n)}_R 
\big]
+\mathrm{h.c.}\right\}.
\end{equation}
These terms, together with the gauge couplings and KK masses allow us
to compute the physical masses and physical couplings to the EW gauge
and would-be Goldstone bosons. These are the required ingredients to
compute the one loop
fermionic contribution to the $T$ parameter and the $Z b
\bar{b}$ coupling. We use the calculation of these two observables
presented in~\cite{Anastasiou:2009rv}, which extended previous
calculations~\cite{Lavoura:1992np} 
to a general enough set-up able to accommodate our fermionic spectrum.  

We have investigated the dependence of the $T$ parameter and $Z b \bar{b}$
coupling as a function of our input parameters, $c_{0,1}^{q,t}$, with
$\lambda_5$ adjusted so that the physical top mass is $m_t=172$ GeV. 
Before discussing the results, we should make two important
comments. First, we are using as an example a very minimal fermionic
spectrum, with no particularly light fermion KK modes. This is enough
to prove that models of EWSB with a soft wall are compatible with EW
precision tests for relatively light KK gauge boson modes. However,
one could expect a richer behavior, similar to the one observed in
hard wall models, if a more complicated fermionic spectrum, including
twisted bc and bulk mixing terms is considered. Second, we have
mentioned above that heavy KK modes propagate deeper in the IR and
therefore they couple more strongly to the Higgs.
This results in a slower decoupling of heavy modes regarding their
contribution to the $T$ parameter (the $Z b \bar{b}$ coupling is less
sensitive to this effect). In fact, we have
observed that the $T$ parameter is quite unstable against the addition
of new fermion KK modes for the first few modes and it stabilizes into
a fixed value only after the inclusion of a relatively large $\sim 10$
number of KK levels. Note that, because of the scaling $m_n \sim
\sqrt{n}$, the addition of many modes does not mean that we have to go
to very high scales before the $T$ parameter is stable. This effect
worsens  the more towards the IR the Higgs is localized (\textit{i.e.}
the larger $\alpha$ is). For values $\alpha \gtrsim 2$, the number of
modes required to stabilize the $T$ parameter gets close to the limit
of strong coupling and it is difficult to make precise quantitative
predictions. Thus, we will show most of our results for $\alpha=1.5$,
although we have checked that these results do not change qualitatively
when $\alpha$ gets closer to $2$.

\begin{figure}[t]
\includegraphics[width=.65\textwidth]{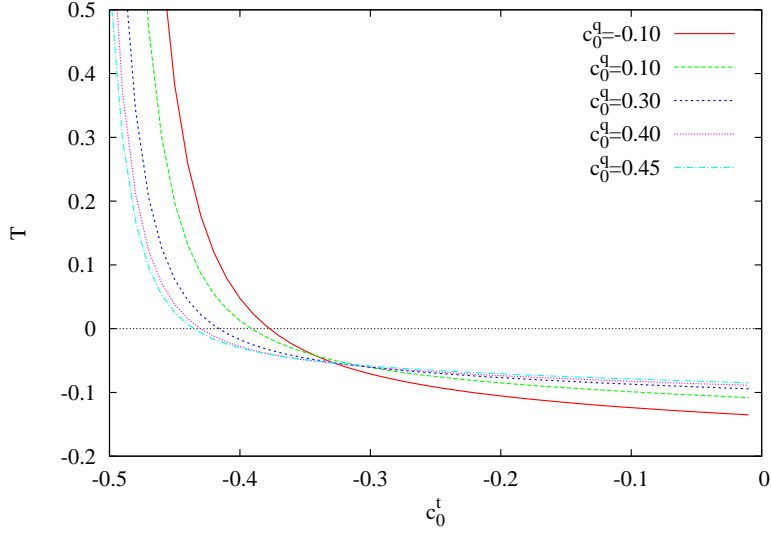}
\caption{$T$ as a function of 
$c_0^t$ for different values of $c_0^q$ with fixed  
$1/L_1=1.2$ TeV,
 $c_1^q=1.2$, $c_1^t=-1$ and $\alpha=1.5$.
\label{Tvsc0t}}
\end{figure}

The results we have obtained for the $T$ parameter and the $Z b
\bar{b}$ coupling can be summarized as follows.
\begin{itemize}
\item  The \textbf{$T$ parameter} is negative in a large
portion of parameter space, as was already observed in hard-wall
models with the same fermionic quantum
numbers~\cite{Carena:2006bn}. It is negative, with a very mild dependence on
$c_0^{q,t}$ 
for values $c_0^t \gtrsim -0.35$, whereas it develops a strong
dependence on the two parameters $c_0^{q,t}$ for values of $c_0^t
\gtrsim 0.35$ becoming quickly positive and of order one. 
This behavior can be clearly seen in Fig.~\ref{Tvsc0t} where we
plot the $T$ parameter as a function of $c_0^t$ for different values
of $c_0^q$. The behavior shown extends to positive values of $c_0^t$
without any significant change.
\item The \textbf{$Zb\bar{b}$ coupling} is much less sensitive to the
  different localization parameters, $c_{0,1}^{q,t}$. In fact, it is
  not even very sensitive to the particular value of $L_1$ (provided
  it is of $\sim$ TeV$^{-1}$ size) for a fixed value of the Higgs
  profile. The reason is that, increasing $L_1^{-1}$, which makes the
  KK modes more massive and therefore should decouple them, also makes
  the Higgs more IR localized, which increases the required value of
  $\lambda_5$ and the coupling among the heavy KK modes in such a way
  that both effects almost cancel each other giving rise to an
  essentially constant value of the $Z b \bar{b}$ coupling $\delta
  g_{b_L} \approx -(1-1.5)\times 10^{-3}$. This effect is in principle
  also present in the $T$ parameter but it is overcome by 
  the larger number of heavy modes that contribute in that case and
  the strong dependence on $c_0^{q,t}$.
\end{itemize}

\subsection{Fit to electroweak observables}

\begin{figure}[t]
\includegraphics[width=.65\textwidth]{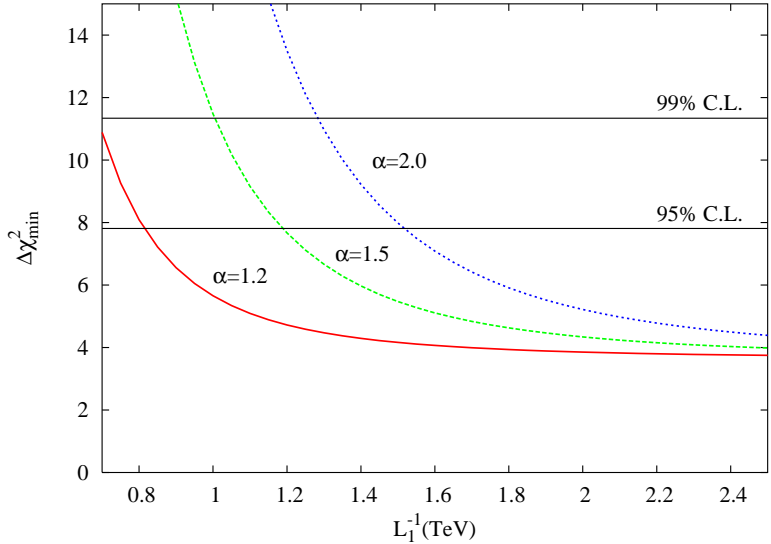}
\caption{
Minimum value of $\Delta \chi^2$ as described in the text as a
function of $L_1^{-1}$ (in TeV) for different values of the Higgs
profile. The horizontal lines show the $95\%$ and $99\%$ C.L. limits
on the $\Delta \chi^2$ for a fit with three variables.
\label{chi2vsL1}}
\end{figure}

Our analysis of the contribution to EW precision observables from the
fermionic sector showed two main features for the class of models we
have considered: the $Z b \bar{b}$ coupling
receives a non-negligible, almost constant correction $\delta g_{b_L}
\approx -(1 - 1.5) \times 10^{-3}$, and the $T$ parameter can get
essentially any value for $c_0^t \lesssim -0.35$. Using these features
we have performed a three parameter fit to all relevant electroweak
precision observables, using an updated version of the code
in~\cite{Han:2004az} (see~\cite{Anastasiou:2009rv} for details on the
fit). The result is summarized in Fig.~\ref{chi2vsL1} in which we show
the minimum value of the $\Delta \chi^2$ obtained in our model
as a function of $L_1^{-1}$
for different values of the Higgs profile. The $\Delta \chi^2$ is 
defined as the $\chi^2$ for an arbitrary point of our model with 
the minimum $\chi^2$ optimizing the values of
$T$, $S$ and $Z b \bar{b}$ coupling subtracted. In the figure we also
show the values of $\Delta \chi^2$ that correspond to 
$95\%$ and $99\%$ C.L. limits for a fit to three variables.
The corresponding bounds on $L_1^{-1}$ read, at $95\%$ C.L.
\begin{equation}
L_1^{-1} \mbox{ (in TeV)} \lesssim 0.85,~1.2,~1.55 \mbox{ for
}\alpha=1.2~,1.5~2,
\end{equation}
which translate into the following values for the gauge boson KK modes
\begin{equation}
m^{\mathrm{GB}}_n \sim \left\{ 
\begin{array}{ll}
1.7,~2.4,~3,\ldots & (\alpha=1.2), \\
2.4,~3.4,~4.2,\ldots & (\alpha=1.5), \\
3.1,~4.4,~5.4,\ldots & (\alpha=2). 
\end{array}\right.
\end{equation}
Note that, even for $\alpha$ close to one, the bound is more stringent
than one would naively get from the $S$ parameter alone. The main
reason is the almost constant contribution to the $Z b \bar{b}$
coupling that creates some tension in the fit. Of course, a richer
fermionic spectrum than we have considered in our minimal model could
result in a different value of this coupling and therefore a smaller bound.

\section{Discussion \label{conclusions}}

Models with warped extra dimensions and a soft IR wall represent a
more general approach to EWSB than hard wall
models. The spectrum of KK excitations is sensitive to the details of
the soft wall and an analysis of the bosonic sector, assuming the SM
fermions to be localized at the UV brane, shows that current
EW constraints are compatible with very light $\sim$ TeV new
resonances. UV localized fermions are a good approximation regarding
the implications on EW observables of the light
fermions. However, 
flavor constraints and top dependent contributions to EW observables
cannot be studied in that approximation. We have developed
the tools to analyze bulk fermions in soft wall models with great 
generality. By assuming a position dependent Dirac mass for the bulk
fermions, which could be generated by a direct coupling to the
soft wall, we can perform the KK expansion of bulk fermions in a very
general set-up. Our construction reproduces well the most relevant
features of bulk fermions present in hard wall models, like the effect
of non-trivial IR boundary conditions, the presence of ultra-light
modes for twisted boundary conditions, flavor universality of
couplings of UV localized fermions to KK gauge bosons or hierarchical
Yukawa couplings through wave function localization.
Using these techniques we have studied the flavor structure of
realistic models with the result that a similar flavor protection as the one
observed in hard wall models can be expected. Similarly, we have
computed the contribution of the top sector to EW precision
observables and shown that simple realistic models with custodial
symmetry and KK excitations as light as
\begin{equation}
m_1 \gtrsim 1.7~\mbox{TeV},
\end{equation} 
can be compatible with all EW precision tests. Furthermore,
the particular realization of the soft wall we have considered
predicts a linear scaling of the mass \textit{squared} of the KK
excitations. As an example, for certain Higgs profile ($\alpha=1.5$)
we can have, at the $95\%$ C.L., the following spectrum of new gauge
boson KK masses
\begin{equation}
m_n \approx 2 \sqrt{n}\times  1.2~\mbox{TeV} \approx
(2.4,~3.2,~4.2)~\mbox{TeV}. 
\end{equation}
Detailed analyses in hard wall models showed that masses 
$m_{KK}^{\mathrm{G}}\lesssim 4-5$ TeV in the case of KK
gluons~\cite{Agashe:2006hk} 
and $m_{KK}^{\mathrm{EW}}
\lesssim 2$ TeV in the case of KK excitations of the EW
bosons~\cite{Agashe:2007ki} could be reached at the LHC
with $\sim 100-300~\mathrm{fb}^{-1}$. It would therefore seem that the
first and maybe even the second mode of KK gluons could be observable
at the LHC although a detailed analysis, taking into account the
details of the spectrum in our model, is required to fully assess the
LHC reach. 

Regarding the fermionic sector, we have chosen a  very minimal set-up,
with just the minimal number of five-dimensional fields to reproduce
the observed spectrum. In that case no particularly
light KK fermions are present in the spectrum. Due to the boundary
conditions, the KK excitations of $X$, which from the SM point of view
is an $SU(2)_L$ doublet of hypercharge $7/6$  are among the lightest
new fermions with a mass $m_1^X \sim 2-2.5$
TeV. This multiplet includes a charge $5/3$ quark that 
was shown to be easily reachable in an early LHC run through pair
production provided it is
light enough~\cite{Contino:2008hi}. The analysis in that reference
only considered $0.5$ and $1$ TeV masses. Most likely,
masses as heavy as the ones we obtain are not reachable at the LHC
through pair production. Single production is a more likely
possibility but again a detailed analysis would be required to
understand the real reach.
This simple fermionic spectrum was chosen for simplicity. It is not a
constraint coming from the soft wall. Richer structures are easy to
implement. For instance, one could simulate the spectrum used in
hard wall realistic models of gauge-Higgs 
unification~\cite{Carena:2006bn,Panico:2008bx} by introducing further bulk
fermions with twisted boundary conditions and a position dependent
mass term mixing them. In that case we can only solve analytically for
a common $z$-independent mass, which is still tolerable for the top
quark. This richer fermionic spectrum would lead to lighter quarks
associated to the top and could make new regions in parameter space
compatible with EW precision data through their contribution
to EW precision observables. Bulk fermions can be also
implemented in actual composite Higgs models in the
soft wall~\cite{Falkowski:2008fz} using the techniques developed in
this work. 

Our results show that soft wall models have the potential to become
realistic models of EWSB readily accessible
at the LHC. We have gone a step forward towards this goal by including
with great generality bulk fermions in these models. However, several
further steps need to be taken before we can consider these models a
full satisfactory solution to the hierarchy problem. One of the most
pressing open questions is that of the stability of the soft wall and
the corresponding $L_1/L_0 \sim \mathrm{TeV}/M_{\mathrm{Pl}}$
hierarchy. Also, given the new bounds and the different patterns of
masses and mixings that can be obtained in soft wall models, a new
analysis of the LHC reach would be welcome. Finally, soft wall models
open the possibility to study a priori completely different physics,
like unparticles, and their relation to EWSB. It has been recently
shown that the Higgs itself could be part of the conformal sector
(Unhiggs). It can be modeled by a continuum of resonances in a
5D soft wall model~\cite{Stancato:2008mp,Falkowski:2008yr}, being
(partly) responsible for EWSB and longitudinal gauge boson scattering
and even being able to reproduce the quantum contributions of a
standard Higgs to the oblique parameters despite having modified
(suppressed) couplings to the SM fields. It would be very interesting
to use the formalism we have developed here to analyze the
effect of the top quark propagating in such background.

\begin{acknowledgments}
This work was supported by
the Swiss National Science Foundation under contract 200021-117873.
\end{acknowledgments}

\appendix

\section{Background solution \label{background}}

The soft wall model that we have considered in this article can be
obtained dynamically from a five dimensional gravitational
model~\cite{Batell:2008zm,Batell:2008me}. We collect the relevant
results in this appendix. More details can be found in the original
references.  
The action describes gravity coupled to two scalars,
the dilaton $\Phi$ and the tachyon $T$,
\begin{eqnarray}
S &=& \int
d^5x\sqrt{g}\,
\Bigg[M^3R-\frac{1}{2}g^{MN}
\partial_M\Phi\partial_N\Phi-\frac{1}{2}g^{MN}\partial_M
T\partial_N T-V(\Phi,T)\Bigg]\nonumber\\ 
&\ &\hspace{20mm} -\int d^4x\sqrt{-g_{UV}}\lambda_{UV}(\Phi,T)\,,
\end{eqnarray}
where $M$ is the $5$D Planck mass and $V(\Phi,T)$ and
$\lambda_{UV}(\Phi,T)$ are the scalar bulk and UV boundary potentials
for the dilaton and the tachyon and are given by 
\begin{eqnarray}
V(\Phi,T)&=&
18\Bigg[\left(\frac{\partial W}{\partial \Phi}\right)
+\left(\frac{\partial W}{\partial T}\right)\Bigg]
-\frac{12}{M^3}W^2\,,\nonumber\\
\lambda_{UV}(\Phi,T)&=&6\Bigg[W(\Phi_0,T_0)
+\partial_{\Phi}W(\Phi_0,T_0)(\Phi-\Phi_0)
+\partial_TW(\Phi_0,T_0)(T-T_0)+\dots\Bigg]\,,
\end{eqnarray}
where $W$ is 
\begin{equation}
W(\Phi,T)=\frac{M^3}{L_0}\Bigg[(\nu-1){\rm e}^{T^2/(24(1+1/\nu)M^3)}-\nu\left(1-\frac{\Phi}{\sqrt{6}M^{3/2}}\right){\rm e}^{\Phi/(\sqrt{6}M^{3/2})}\Bigg]\,.
\end{equation}
Solutions for the equations of motions for the dilaton and the tachyon using the above potentials can be found as
\begin{eqnarray}
\Phi(z)&=&\sqrt{\frac{8}{3}}M^{3/2}\left(\frac{z}{L_1}\right)^{\nu}\,,\nonumber\\
T(z)&=&\pm 4\sqrt{1+1/\nu}M^{3/2}\left(\frac{z}{L_1}\right)^{\nu/2}\,,
\end{eqnarray}
where the background metric is chosen to be
\begin{equation}
g_{MN}=a^2(z)\,{\rm e}^{-4/3(z/L_1)^{\nu}}\, \eta_{MN}.
\end{equation}
Note that one can recover the metric in Eq.~(\ref{metric}) and the action in Eq.~(\ref{action}) by the redefinitions
\begin{eqnarray}
\Phi &\rightarrow& \sqrt{\frac{3}{8}}M^{-3/2}\Phi\,\nonumber\\
g_{MN} &\rightarrow& {\rm e}^{4/3(z/L_1)^{\nu}} g_{MN}\,.
\end{eqnarray}
Throughout this paper, we considered the case with $\nu=2$ only.

\section{Bulk Bosonic Fields in the Soft Wall \label{bosons}}
In this Appendix we review the Kaluza-Klein expansion of bulk bosonic
fields in the soft wall. Further details can be found in~\cite{Batell:2008me}.

\subsection{Bulk Higgs}

Let us assume the following form for the bulk and the UV boundary Higgs
potentials $V(H)$ and $V_{UV}(H)$ in Eq.~(\ref{bosonicL}) 
\begin{eqnarray}
V(H)&=&m_H^2(z)\,{\rm Tr}\,|H|^2\,,\nonumber\\
V_{UV}(H)&=&\lambda_0\,L_0^2 \left( {\rm Tr}\,|H|^2-v_0^2\right)^2\,,
\end{eqnarray}
where the effective mass for the bulk Higgs is defined to have the form
\begin{equation}
m_H(z)^2=\frac{1}{L_0^2}\left[\alpha(\alpha-4)-2\alpha
\frac{z^2}{L_1^2}\right]\,. 
\end{equation}
The above mass term is assumed to arise from a coupling to another
scalar which gets a background vev. The UV boundary potential is
added to the action in Eq.~(\ref{bosonicL}) so that the solutions of
the equations of motion for the Higgs field satisfy non-trivially the
UV boundary conditions. 
Solving for the equation of motion for the
bulk Higgs using $V(H)$ and demanding that the solution is finite in
the soft wall background, one finds 
\begin{equation}
f_h(z)=c_h\, z^{\alpha}\,,
\end{equation}
where $c_h$ is a normalization constant and 
we have defined the Higgs vev as
\begin{equation}
\langle H \rangle = \frac{f_h(z)}{\sqrt{2}} \begin{pmatrix} v & 0 \\ 0 &
  v \end{pmatrix},
\end{equation}
with $v$ a constant with mass dimension 1.
The UV boundary condition satisfied by the above solution is given in
Ref.~\cite{Batell:2008me}. Properly normalizing the bulk Higgs field
and solving for the mass of the $W$ boson in terms of the overlaps of
the bulk Higgs profile with the gauge field zero modes for the $W$ we
find that 
\begin{equation}
c_h=\sqrt{\frac{2
    L_1^2}{L_0^3}}\frac{1}{\sqrt{\Gamma(\alpha-1,L_0^2/L_1^2)}}\, 
\end{equation}
with $\Gamma$ the incomplete Gamma function, 
and $v=174$ GeV, up to corrections $\mathcal{O}(v^2 L_1^2)$.
It was shown in~\cite{Batell:2008me} that this solution is indeed the
ground state of the theory.

\subsection{Gauge bosons} 

The expansion of gauge bosons in our background was also considered
in~\cite{Batell:2008me} and we collect here the main results.
The action for a bulk $U(1)$ gauge field reads
\begin{equation}
S=-\frac{1}{4}\,\int d^5x\,\sqrt{g}{\rm e}^{-\Phi}\,A_{MN}A^{MN}\,.
\end{equation}
We perform the standard KK decomposition
\begin{equation}
A_{\mu}(x,z)=\sum_{n}f_n^A(z) A_{\mu}^{(n)}(x),
\end{equation}
where the four dimensional fields $A_{\mu}^{(n)}(x)$ satisfy the four
dimensional equations of motion for a gauge massive gauge boson with
mass $m_n$.
The profile functions $f_n^A(z)$ satisfy then the following equation 
\begin{equation}
\Big[
\partial_5^2 + \left (\frac{a^\prime}{a} - \Phi^\prime \right)
\partial_5 + m_n^2 \Big] f_n^A = 0,
\end{equation}
and normalization condition
\begin{equation}
\int_{L_0}^\infty dz\, a e^{-\Phi} f_m^A f_n^A =
\delta_{mn}. \label{orthonormality:gauge} 
\end{equation}
Applying the change of variable 
\begin{equation}
x=\frac{z^2}{L_1^2}\,,
\end{equation}
and inserting the explicit form of the
metric and the dilaton field,
the equation for the bosonic profile reads
\begin{equation}
\Bigg[x\partial^2_x-x\partial_x+\frac{L_1^2\,m_n^2}{4}\Bigg]f_A^n(z)=0\,,
\end{equation}
which is the same equation $g(z)$ of Eq.~(\ref{fermion:change})
satisfied with the special values of fermionic parameters $c_0=1/2$
and $c_1=1$. The normalizable solution is given in terms of confluent
hypergeometric function
\begin{equation}
f_n^A(z)= N_n^A\,U\left(\frac{-m_n^2\,L_1^2}{4},0,\frac{z^2}{L_1^2}\right)\,,
\end{equation}
where the normalization constants $N_n^A$ 
are determined from Eq.~(\ref{orthonormality:gauge}). 
The masses for the massive gauge bosons are found by applying the
corresponding boundary conditions. Applying Neumann boundary
conditions for the UV brane one finds that the masses for the heavy
modes are given approximately by 
\begin{equation}
m_n^2\sim \frac{4}{L_1^2}\,n^2\,.
\end{equation}
$U(0,0,z^2)=1$ so the profile for a massless gauge boson zero mode
is given entirely by its normalization constant, 
\begin{equation}
f_A^0(z)=\sqrt{\frac{2}{L_0\,E_1(L_0^2/L_1^2)}}\,,
\end{equation}
where $E_\nu(z)= \int_1^\infty dt\, e^{-zt}/t^\nu$ is the Exponential
Integral E function.


\begin{thebibliography}{99}

\bibitem{Randall:1999vf}
  L.~Randall and R.~Sundrum,
  Phys.\ Rev.\ Lett.\  {\bf 83} (1999) 4690
  [hep-th/9906064];
  Phys.\ Rev.\ Lett.\  {\bf 83} (1999) 3370
  [hep-ph/9905221].


\bibitem{Maldacena:1997re}
  J.~M.~Maldacena,
  Adv.\ Theor.\ Math.\ Phys.\  {\bf 2} (1998) 231
  [Int.\ J.\ Theor.\ Phys.\  {\bf 38} (1999) 1113]
  [hep-th/9711200];
  S.~S.~Gubser, I.~R.~Klebanov and A.~M.~Polyakov,
  Phys.\ Lett.\  B {\bf 428} (1998) 105
  [hep-th/9802109];
  E.~Witten,
  Adv.\ Theor.\ Math.\ Phys.\  {\bf 2} (1998) 253
  [hep-th/9802150].

\bibitem{ArkaniHamed:2000ds}
  N.~Arkani-Hamed, M.~Porrati and L.~Randall,
  JHEP {\bf 0108} (2001) 017
  [hep-th/0012148];
  R.~Rattazzi and A.~Zaffaroni,
  JHEP {\bf 0104} (2001) 021
  [hep-th/0012248];
  M.~Perez-Victoria,
  JHEP {\bf 0105} (2001) 064
  [hep-th/0105048].


\bibitem{Carena:2006bn}
  M.~S.~Carena, E.~Ponton, J.~Santiago and C.~E.~M.~Wagner,
  Nucl.\ Phys.\  B {\bf 759} (2006) 202
  [hep-ph/0607106];
  Phys.\ Rev.\  D {\bf 76} (2007) 035006
  [hep-ph/0701055].
  R.~Contino, L.~Da Rold and A.~Pomarol,
  Phys.\ Rev.\  D {\bf 75} (2007) 055014
  [hep-ph/0612048].

\bibitem{Cacciapaglia:2006gp}
  G.~Cacciapaglia, C.~Csaki, G.~Marandella and J.~Terning,
  Phys.\ Rev.\  D {\bf 75} (2007) 015003
  [hep-ph/0607146].

\bibitem{Agashe:2003zs}
  K.~Agashe, A.~Delgado, M.~J.~May and R.~Sundrum,
  JHEP {\bf 0308} (2003) 050
  [hep-ph/0308036].

\bibitem{Agashe:2006at}
  K.~Agashe, R.~Contino, L.~Da Rold and A.~Pomarol,
  Phys.\ Lett.\  B {\bf 641} (2006) 62
  [hep-ph/0605341].

\bibitem{Davoudiasl:2002ua}
  H.~Davoudiasl, J.~L.~Hewett and T.~G.~Rizzo,
  Phys.\ Rev.\  D {\bf 68} (2003) 045002
  [hep-ph/0212279];
  M.~S.~Carena, A.~Delgado, E.~Ponton, T.~M.~P.~Tait and C.~E.~M.~Wagner,
  Phys.\ Rev.\  D {\bf 68} (2003) 035010
  [hep-ph/0305188];
  Phys.\ Rev.\  D {\bf 71} (2005) 015010
  [hep-ph/0410344];
  A.~Djouadi, G.~Moreau and F.~Richard,
  Nucl.\ Phys.\  B {\bf 773} (2007) 43
  [hep-ph/0610173];
  C.~Bouchart and G.~Moreau,
  Nucl.\ Phys.\  B {\bf 810} (2009) 66
  [0807.4461 [hep-ph]].

\bibitem{Karch:2006pv}
  A.~Karch, E.~Katz, D.~T.~Son and M.~A.~Stephanov,
  Phys.\ Rev.\  D {\bf 74} (2006) 015005
  [hep-ph/0602229].

\bibitem{Falkowski:2008fz}
  A.~Falkowski and M.~Perez-Victoria,
  JHEP {\bf 0812} (2008) 107
  [0806.1737 [hep-ph]].

\bibitem{Batell:2008me}
  B.~Batell, T.~Gherghetta and D.~Sword,
  Phys.\ Rev.\  D {\bf 78} (2008) 116011
  [0808.3977 [hep-ph]].

\bibitem{Strassler:2006im}
  M.~J.~Strassler and K.~M.~Zurek,
  Phys.\ Lett.\  B {\bf 651} (2007) 374
  [hep-ph/0604261].

\bibitem{Georgi:2007ek}
  H.~Georgi,
  Phys.\ Rev.\ Lett.\  {\bf 98} (2007) 221601
  [hep-ph/0703260].

\bibitem{Cacciapaglia:2008ns}
  G.~Cacciapaglia, G.~Marandella and J.~Terning,
  JHEP {\bf 0902} (2009) 049
  [0804.0424 [hep-ph]].

\bibitem{Falkowski:2008yr}
  A.~Falkowski and M.~Perez-Victoria,
  0810.4940 [hep-ph];
 A.~Falkowski and M.~Perez-Victoria,
  0901.3777 [hep-ph].

\bibitem{Shiu:2007tn}
  G.~Shiu, B.~Underwood, K.~M.~Zurek and D.~G.~E.~Walker,
  Phys.\ Rev.\ Lett.\  {\bf 100} (2008) 031601
  [0705.4097 [hep-ph]];
  P.~McGuirk, G.~Shiu and K.~M.~Zurek,
  JHEP {\bf 0803} (2008) 012
  [0712.2264 [hep-ph]].


\bibitem{Delgado:2009xb}
  A.~Delgado and D.~Diego,
  0905.1095 [hep-ph].

\bibitem{Abramowitz}
     M.~Abramowitz and I.~A.~Stegun (Ed.)",
     "Handbook of Mathematical Functions, Dec. 1972".

\bibitem{DelAguila:2001pu}
  F.~Del Aguila and J.~Santiago,
  JHEP {\bf 0203} (2002) 010
  [hep-ph/0111047];
  K.~Agashe and G.~Servant,
  Phys.\ Rev.\ Lett.\  {\bf 93}, 231805 (2004)
  [hep-ph/0403143];
  JCAP {\bf 0502}, 002 (2005)
  [hep-ph/0411254].

\bibitem{flavor}
  Y.~Grossman and M.~Neubert,
  Phys.\ Lett.\  B {\bf 474}, 361 (2000)
  [hep-ph/9912408];
  T.~Gherghetta and A.~Pomarol,
  Nucl.\ Phys.\  B {\bf 586}, 141 (2000)
  [hep-ph/0003129];
  S.~J.~Huber and Q.~Shafi,
  Phys.\ Lett.\  B {\bf 498}, 256 (2001)
  [hep-ph/0010195];
  G.~Burdman,
  Phys.\ Rev.\  D {\bf 66}, 076003 (2002)
  [hep-ph/0205329];
  Phys.\ Lett.\  B {\bf 590}, 86 (2004)
  [hep-ph/0310144];
  S.~J.~Huber,
  Nucl.\ Phys.\  B {\bf 666}, 269 (2003)
  [hep-ph/0303183];
  K.~Agashe, G.~Perez and A.~Soni,
  Phys.\ Rev.\ Lett.\  {\bf 93}, 201804 (2004)
  [hep-ph/0406101];
  Phys.\ Rev.\  D {\bf 71}, 016002 (2005)
  [hep-ph/0408134];
  G.~Cacciapaglia, C.~Csaki, J.~Galloway, G.~Marandella, J.~Terning
  and A.~Weiler, 
  JHEP {\bf 0804}, 006 (2008)
  [0709.1714 [hep-ph]];
  A.~L.~Fitzpatrick, G.~Perez and L.~Randall,
  0710.1869 [hep-ph];
  C.~Csaki, A.~Falkowski and A.~Weiler,
  JHEP {\bf 0809}, 008 (2008)
  [0804.1954 [hep-ph]];
  J.~Santiago,
  JHEP {\bf 0812}, 046 (2008)
  [0806.1230 [hep-ph]];
  C.~Csaki, A.~Falkowski and A.~Weiler,
  0806.3757 [hep-ph];
  S.~Casagrande, F.~Goertz, U.~Haisch, M.~Neubert and T.~Pfoh,
  JHEP {\bf 0810}, 094 (2008)
  [0807.4937 [hep-ph]];
  M.~Blanke, A.~J.~Buras, B.~Duling, S.~Gori and A.~Weiler,
  JHEP {\bf 0903}, 001 (2009)
  [0809.1073 [hep-ph]];
  K.~Agashe, A.~Azatov and L.~Zhu,
  0810.1016 [hep-ph].
  M.~Blanke, A.~J.~Buras, B.~Duling, K.~Gemmler and S.~Gori,
  JHEP {\bf 0903}, 108 (2009)
  [0812.3803 [hep-ph]];
  K.~Agashe,
  0902.2400 [hep-ph].
  M.~E.~Albrecht, M.~Blanke, A.~J.~Buras, B.~Duling and K.~Gemmler,
  0903.2415 [hep-ph];



\bibitem{Csaki:2009bb}
  C.~Csaki and D.~Curtin,
  0904.2137 [hep-ph].

\bibitem{Goertz:2008vr}
  F.~Goertz and T.~Pfoh,
  JHEP {\bf 0810} (2008) 035
  [0809.1378 [hep-ph]].

\bibitem{Barbieri:2004qk}
  R.~Barbieri, A.~Pomarol, R.~Rattazzi and A.~Strumia,
  Nucl.\ Phys.\  B {\bf 703} (2004) 127
  [hep-ph/0405040].


\bibitem{Barbieri:2003pr}
  R.~Barbieri, A.~Pomarol and R.~Rattazzi,
  Phys.\ Lett.\  B {\bf 591} (2004) 141
  [hep-ph/0310285].

\bibitem{Anastasiou:2009rv}
  C.~Anastasiou, E.~Furlan and J.~Santiago,
  0901.2117 [hep-ph].

\bibitem{Lavoura:1992np}
  L.~Lavoura and J.~P.~Silva,
  Phys.\ Rev.\  D {\bf 47} (1993) 2046;
  P.~Bamert, C.~P.~Burgess, J.~M.~Cline, D.~London and E.~Nardi,
  Phys.\ Rev.\  D {\bf 54} (1996) 4275
  [hep-ph/9602438].

\bibitem{Han:2004az}
  Z.~Han and W.~Skiba,
  Phys.\ Rev.\  D {\bf 71} (2005) 075009
  [hep-ph/0412166];
  Z.~Han,
  Phys.\ Rev.\  D {\bf 73} (2006) 015005
  [hep-ph/0510125].

\bibitem{Agashe:2006hk}
  K.~Agashe, A.~Belyaev, T.~Krupovnickas, G.~Perez and J.~Virzi,
  Phys.\ Rev.\  D {\bf 77} (2008) 015003
  [hep-ph/0612015];
  B.~Lillie, L.~Randall and L.~T.~Wang,
  JHEP {\bf 0709} (2007) 074
  [hep-ph/0701166].

\bibitem{Agashe:2007ki}
  K.~Agashe {\it et al.},
  Phys.\ Rev.\  D {\bf 76} (2007) 115015
  [0709.0007 [hep-ph]];
  K.~Agashe, S.~Gopalakrishna, T.~Han, G.~Y.~Huang and A.~Soni,
  0810.1497 [hep-ph].

\bibitem{Contino:2008hi}
  R.~Contino and G.~Servant,
  JHEP {\bf 0806} (2008) 026
  [0801.1679 [hep-ph]].


\bibitem{Panico:2008bx}
  G.~Panico, E.~Ponton, J.~Santiago and M.~Serone,
  Phys.\ Rev.\  D {\bf 77} (2008) 115012
  [0801.1645 [hep-ph]];
  M.~Carena, A.~D.~Medina, N.~R.~Shah and C.~E.~M.~Wagner,
  0901.0609 [hep-ph].

\bibitem{Stancato:2008mp}
  D.~Stancato and J.~Terning,
  0807.3961 [hep-ph].

\bibitem{Batell:2008zm}
  B.~Batell and T.~Gherghetta,
  Phys.\ Rev.\  D {\bf 78} (2008) 026002
  [0801.4383 [hep-ph]].


\end{thebibliography}
\end{document}